\documentclass[twocolumn,aps]{revtex4}
\usepackage{epsfig}
\usepackage{times}

\begin{document}

\newcommand{\RRG}[1][ ]{RRG#1}

\title{Evolutionary prisoner's dilemma games with optional participation}
\author{Gy\"orgy Szab\'o}
\affiliation
{Research Institute for Technical Physics and Materials Science
P.O.Box 49, H-1525 Budapest, Hungary}
\author{Christoph Hauert}
\affiliation{Department of Zoology, University of British
Columbia, 6270 University Boulevard, Vancouver, B.C., Canada V6T 1Z4}

\date{\today}

\begin{abstract}
Competition among cooperators, defectors, and loners is studied in an
evolutionary prisoner's dilemma game with optional participation. Loners are
risk averse i.e. unwilling to participate and rather rely on small but fixed
earnings. This results in a rock-scissors-paper type cyclic dominance of the
three strategies. The players are located either on square lattices or random
regular graphs with the same connectivity. Occasionally, every player
reassesses its strategy by sampling the payoffs in its neighborhood. The 
loner strategy efficiently prevents successful spreading of selfish,
defective behavior and avoids deadlocks in states of mutual defection.
On square lattices, Monte Carlo simulations reveal self-organizing patterns
driven by the cyclic dominance, whereas on random regular graphs different
types of oscillatory behavior are observed: the temptation to defect 
determines whether damped, periodic or increasing oscillations occur.
These results are compared to predictions by pair approximation. Although
pair approximation is incapable of distinguishing the two scenarios because
of the equal connectivity, the average frequencies as well as the 
oscillations on random regular graphs are well reproduced.
\end{abstract}

\maketitle

\section{\label{sec:intr}Introduction}

Evolutionary prisoner's dilemma games (PDG) 
\cite{hofbauer:98,maynard:82,weibull:95,hamilton:amnat63,axelrod:84} were
introduced to study the emergence and maintenance of cooperation among
selfish individuals in societies where strategies are either inherited or
adopted through basic imitation rules. In the original PDG  \cite{neumann:44}
two players simultaneously decide whether to cooperate or defect. 
Mutual cooperation (defection) yields the highest (lowest) \emph{collective}
payoff which is shared equally. However, still higher \emph{individual} 
payoffs are achieved by defectors facing cooperators leaving the latter with
the lowest possible payoff. Thus, defection is dominant because they fare
better (or at least equal) regardless of the co-players decision.
Consequentially, 'rational' players always end up with the lowest
collective payoff instead of the higher reward for mutual cooperation.

Obviously, this result is at odds with observations in human and animal
societies. Indeed, cooperation may emerge under certain circumstances
(see e.g. 
\cite{trivers:qrevbiol71,nowak:nature98,axelrod:science81,hauert:science02,
nowak:nature92:space,sigmund:pnas01}). 
In well-mixed populations, i.e. with random matchings, cooperative strategies
succeed provided that interactions between the same individuals are repeated
with sufficiently high probability. 
Computer tournaments \cite{axelrod:84,nowak:nature93} emphasized the
importance of particularly simple strategies such as $D$ (always defect),
$C$ (always cooperate), and $T$ (''tit for tat'', cooperate on the first
move and then repeat the co-player's move).

In spatially extended systems with limited local interaction, cooperators
may thrive by forming clusters 
\cite{nowak:nature92:space,hubermann:pnas93,nowak:jbifchaos94} and thereby 
reducing exploitation by defectors. For example, on square lattices with 
$C$, $D$ strategies and suitable parameter values, clusters of cooperators 
typically expand along straight boundaries while being invaded by defectors
along corners and irregular boundaries. The competition between these two
invasion processes maintains persistent co-existence of both strategies.
When tuning the model parameters, universal phase transitions between
states of homogenous defection, co-existence and mutual cooperation are
observed \cite{szabo:pre98,chiappin:pre99}. Very recently the analysis
is extended to diluted lattices \cite{vainstein:pre01} and to adaptive
networks \cite{zimmermann:01}.

The dynamics of such spatial evolutionary PDG becomes increasingly complex
if players can adopt three or more strategies \cite{szabo:01,
killingback:jtb98,brauchli:jtb99}. In particular, the crucial role of the
$T$ strategy is confirmed \cite{nakamaru:jtb97,szabo:pre00:cdt}. For example,
in an externally driven system supplying $C$, i.e. nourishing defectors,
cooperative behavior persists through the cyclic dominance of $D$, $C$, and
$T$ which maintains self-organizing domain structures 
\cite{tainaka:pre94,frachebourg:jpa98,frean:procb01}.

In this work, we consider another PDG with three strategies by allowing
for optional participation. Players can adopt the $C$ and $D$ strategies 
mentioned above or decline participation relying on some small but fixed
source of income. Such risk averse players are termed loners ($L$). 

Voluntary participation in public goods games \cite{hauert:science02,
szabo:prl02} turned out to be an efficient way to prevent and avoid the
tragedy of the commons \cite{hardin:science68}. This system has an inherent
cyclic dominance resulting in periodic oscillations in well-mixed populations
\cite{hauert:jtb02:vpgg} and self-organizing polydomain structures on square
lattices \cite{hauert:vpgg02}. However note that the former case requires
group sizes larger than pairs i.e. excluding the PDG. Pairwise interactions
invariably lead to homogenous states of loners. In sharp contrast, we
demonstrate that cooperative behavior persists on square lattices producing
self-organizing patterns whereas varying types of oscillations occur on
random regular graphs (\RRG[]) emphasizing the importance of topological
characteristics. \RRG[s] represent a natural choice for social networks
relevant in human behavior as well as economic interactions where the
geometrical location hardly matters.

\section{\label{sec:model}The models}
We consider the evolutionary PDG in large populations with $N$ players in
the limit $N \to \infty$ for different geometries: First, we briefly review 
the results for well-mixed populations i.e. the mean-field approximation.
Second, the players are arranged on a square lattice and interact only with
their four nearest neighbors to the north, east, south and west. Finally,
the players are located on a \RRG with fixed connectivity ($z=4$) excluding
double or self-connectivities. Therefore, each player has the same number
of neighbors as its counterpart on the square lattice. Note that locally a 
\RRG is similar to a tree (or Bethe-lattice) because the average loop size 
increases with $N$ \cite{bollobas:95}. Since boundary problems render
Bethe-lattices unsuitable for Monte Carlo (MC) simulations, \RRG[s] serve 
as suitable substitutes. In addition, the analysis of branching annihilating
random walks has justified the validity of pair approximation on \RRG 
\cite{szabo:pre00:rrg}.

Each player follows one of three strategies, namely defection $D$,
cooperation $C$, and loner $L$, denoted by a three-component unit vector:
\begin{equation}
{\bf s}(x)=\left( \matrix{1 \cr 0 \cr 0 \cr}\right)\ ,~
    \left( \matrix{0 \cr 1 \cr 0 \cr}\right)\ ,~
    \left( \matrix{0 \cr 0 \cr 1 \cr}\right)\ .
\end{equation}
The total payoff $m({\bf x})$ for the player located on site $\bf x$ is 
specified by
\begin{equation}
m({\bf x})=\sum_{\delta } {\bf s}^{T}(\bf x){\bf M}\cdot{\bf s}
( {\bf x} + \delta ),
\end{equation}
where ${\bf s}^{T}(\bf x)$ denotes the transpose of ${\bf s}({\bf x})$ and
the summation runs over the 
four nearest neighbors ($\delta $) as defined by the geometry under 
consideration. Following \cite{nowak:nature92:space}, the payoff matrix 
$\bf M$ is given by
\begin{equation}
{\bf M}=\left( \matrix{0 & b & \sigma \cr
                       0 & 1 & \sigma \cr
                       \sigma & \sigma & \sigma \cr}\right),
\end{equation}
where $b$ ($1<b<2$) determines the temptation to defect. Loners and their
co-players invariably obtain the fixed payoff $0<\sigma<1$, i.e. they
perform better than two defectors but are worse off than cooperative pairs. 

Randomly chosen players are allowed to reassess and modify their strategy.
For example, the player at site $\bf x$ adopts the strategy of one of its 
(randomly chosen) neighbors $\bf y$ with a probability
\begin{equation}
\label{eq:update}
W[{\bf s}({\bf x}) \leftarrow {\bf s}({\bf y})] = {1 \over 1 +
 \exp {[(m({\bf x})-m({\bf y}))/K]} },
\end{equation}
where $K$ introduces some noise that occasionally leads to irrational
decisions.

The imitation rule or learning mechanism (\ref{eq:update}) leads to three 
trivial unstable fixed points, namely the homogenous $C$, $D$ and $L$ states.
The highest average payoff ($m_c=4$) is achieved if everybody cooperates but
this state is vulnerable to exploitation. For sufficiently large $b$ the
appearance of a single defector suffices for cooperation to break down.
The system then evolves into a homogenous $D$ state with the lowest average
payoff ($m_d=0$). Loners may in turn invade this state and thereby provide
an escape hatch out of states of mutual defection. Eventually a homogenous
$L$ state is reached with an average payoff $m_l=4\sigma$. Note that
solitary defectors are eliminated by noise. Finally, the loop is closed by
noting that if no one participates in the PDG, cooperators may thrive 
again. However, also note that this requires the presence of at least
two cooperators. This cyclic rock-scissors-paper type dominance determines
the system's dynamics over a wide parameter range.

\section{\label{sec:mfpa}Mean-field and pair approximations}

\begin{figure}
\centerline{\epsfig{file=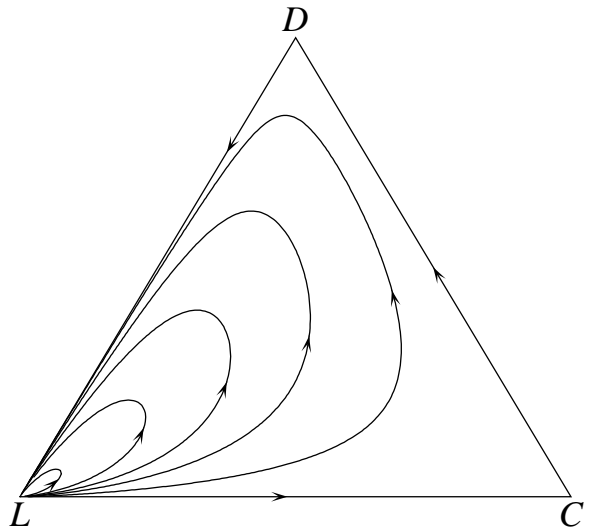,width=8cm}}
\caption{\label{fig:mf}Time evolution of states in mean-field approximation
for $b=1.5$, $K=0.1$ and $\sigma=0.3$. }
\end{figure}

In the classical mean-field approximation, i.e. in well-mixed populations,
this system is described by the (time-dependent) frequencies $f_i(t)$ of
the three strategies with $f_C(t)+f_D(t)+f_L(t)=1$ \cite{szabo:01,
hauert:jtb02:vpgg}. Figure~\ref{fig:mf} displays the time evolution on
a ternary phase diagram.

Regardless of the initial configuration, the system always approaches the
homogenous $L$ state. This result is robust against changes in the parameter
values. The heteroclinic cycle along the boundary of the simplex $S_3$
reflects the cyclic dominance of the strategies. Mean-field approximations
neglect spatial structures, i.e. they describe individuals with an infinite
range of interactions or situations where co-players are randomly chosen.

In contrast, pair approximation is capable of capturing some relevant
effects occurring for lasting (fixed) partnership. The frequency of
strategies is determined by two-site or pair configuration probabilities
(for an introduction see e.g. \cite{marro:99}). The probability of a
strategy change depends on the configuration of an eight-site cluster
consisting of the randomly chosen player, one of its randomly chosen
neighbors (the pair) and their six nearest neighbors. Configuration 
probabilities of larger clusters are assumed to be products of the
corresponding pair configuration probabilities. Due to the same
connectivity $z=4$ of the square lattice and the chosen \RRG[],
the two structures are indistinguishable by pair approximation.

\begin{figure}
\centerline{\epsfig{file=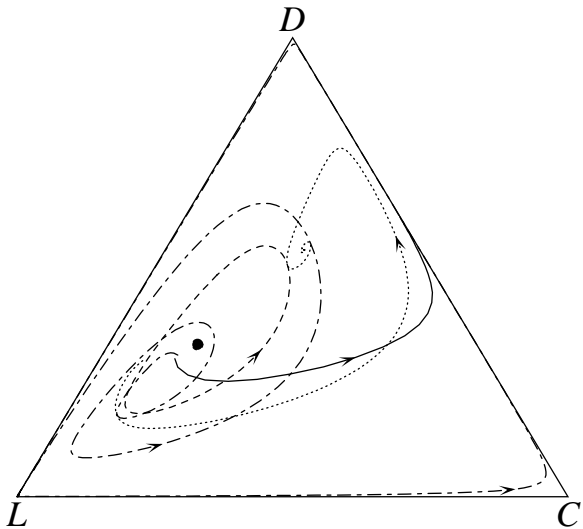,width=8cm}}
\caption{\label{fig:pair}Different trajectories as predicted by pair
approximation: (a) for $b=1.03$ (solid line) loners go extinct and only
$D$ and $C$ survive and co-exist in a stationary state. 
(b) all the three strategies coexist for $b=1.2$ (dotted line). 
(c) for $b=1.5$ the limit cycle is shown (dashed line). The bullet indicates
the unstable interior fixed point.
(d) the trajectory spirals towards the boundary of the simplex ($b=1.9$).}
\end{figure}

The following systematic analysis is restricted to the case of $\sigma = 0.3$
and $K=0.1$. Figure~\ref{fig:pair} shows four typical trajectories for
different values of the temptation to defect $b$. 
For very small $b$ ($1<b<b_0^{(\text{p})}=1.0485(1)$) the clustering advantage
of cooperators suffices to offset exploitation by defectors. Loners are unable
to provide a viable alternative and vanish. The ratio of $D$ and $C$ in the
stationary state depends on the parameters $b$, $\sigma$, and $K$. When
increasing $b$ ($b_0^{(\text{p})} < b < b_1^{(\text{p})}=1.4670(3)$), a
stable interior fixed point appears, i.e. all trajectories approach this
stationary value where all three strategies co-exist. For 
$b > b_1^{(\text{p})}$ periodic oscillations appear. Both, the amplitude
and the period increase with $b$. In finite populations with
$b>b_2^{(\text{p})}\simeq 1.85$ any strategy may go extinct due to noise
and the trajectory ends in one of the absorbing homogenous states. 
Note that in this case the rigorous numerical analysis becomes difficult
because of the extremely low configuration probabilities occurring during
the oscillations.

\section{\label{sec:sre}Monte Carlo simulations}

MC simulations are performed on square lattices and on \RRG[s]. In both cases 
the runs are started from random initial states. Frequencies and payoffs of
all three strategies are traced over time. After an appropriate relaxation
time the average frequencies and payoffs are determined (sampling times
varied from $10^4$ to $10^5$ MC steps per site).

The linear extension of the square lattice with periodic boundary conditions
was $400$ sites for most MC simulations. However, in the vicinity of
transition points the system size was increased to up to $800$ sites in
order to suppress undesired diverging fluctuations.
\begin{figure}
\centerline{\epsfig{file=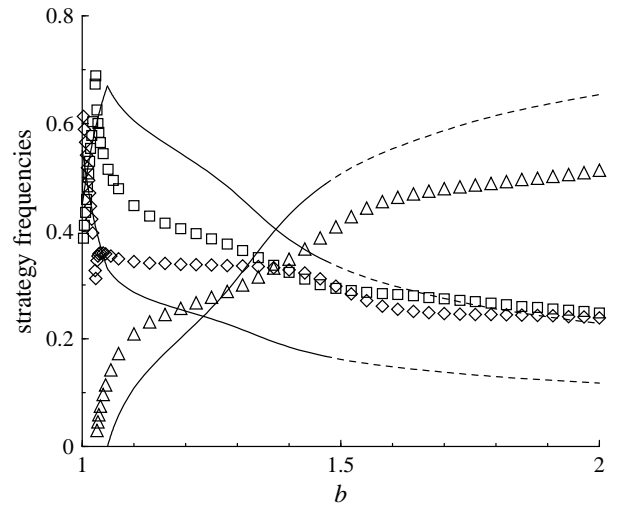,width=8cm}}
\caption{\label{fig:sqmcp}Average frequency of defectors (squares),
cooperators (diamonds), and loners (triangles) as a function of $b$ for
$K=0.1$ and $\sigma=0.3$ on square lattices. The solid (dashed) lines show
the stable (unstable) stationary values predicted by pair approximation.}
\end{figure}

The simulations confirm that loners die out for small
$b<b_0^{(\text{sq})}=1.0262(1)$. In fact, since loners invade the territory
of defectors, they can only survive if the frequency of defectors exceeds
a threshold value. Note that in the vicinity of the transition point,
the pair approximation predicts slightly lower defector frequencies and
therefore a higher threshold value. 

The frequency of loners vanishes continuously when decreasing
$b\to b_0^{(\text{sq})}$. The numerical analysis of MC data suggests that
this extinction process belongs to the directed percolation universality
class \cite{marro:99,hinrichsen:ap00}. Related critical transitions were 
reported and discussed in \cite{szabo:pre98,chiappin:pre99,szabo:prl02}.
In contrast to this non-analytical transition, the pair approximation
predicts a mean-field type behavior, i.e. the frequency of loners vanishes
linearly with $b-b_0^{(\text{p})}$.

For $b>b_0^{(\text{sq})}$ the frequency of loners (defectors) increases
(decreases) monotonously with $b$. However note that the oscillatory
behavior predicted by pair approximation is absent on square lattices.
The time evolution of the lattice shows cyclic invasions for any site but
the phases of these local oscillations are not synchronized
\cite{hauert:vpgg02}. Instead, the cyclic invasion sustains a
self-organizing poly-domain structure. Similar patterns are observed for
spatial rock-scissors-paper games (see e.g.
\cite{szabo:pre00:cdt,tainaka:pre94,frean:procb01}).
\begin{figure}
\centerline{\epsfig{file=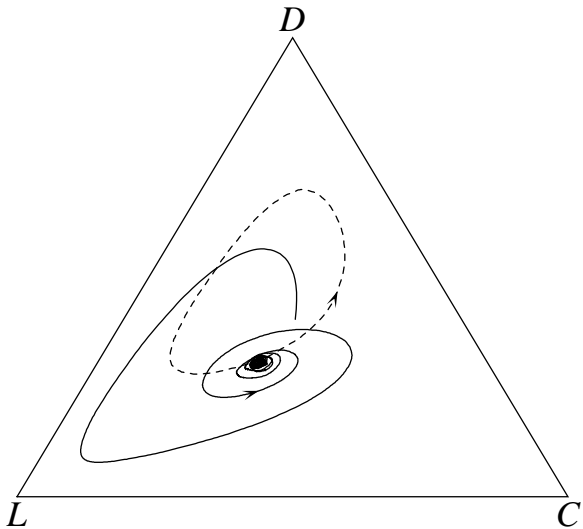,width=8cm}}
\caption{\label{fig:sqgr}Time evolution of strategy frequencies according
to MC simulations for $b=1.5$, $\sigma = 0.3$, and $K=0.1$. The solid line 
shows the trajectory on a square lattice with $600\times 600$ sites.
The dashed line indicates the limit cycle on a \RRG with $N=5 \cdot 10^5$ 
sites.}
\end{figure}

\begin{figure}
\centerline{\epsfig{file=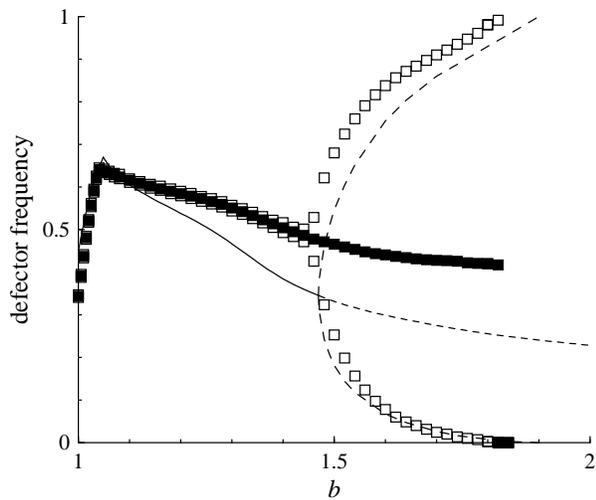,width=8cm}}
\caption{\label{fig:dminmax}Average frequency of defectors versus temptation
$b$ obtained by MC simulations (closed squares) on \RRG and by pair
approximation (solid line). The minimum and maximum frequency of $D$ is
indicated by open symbols (simulations) and dashed lines (pair 
approximation).}
\end{figure}

To illustrate the striking difference between the results obtained on square
lattices and \RRG[], we compare two trajectories determined by MC simulations
for the same model parameters (see Fig.~\ref{fig:sqgr}). On square lattices
the trajectory evolves towards a stable interior fixed point. In contrast,
on \RRG the trajectory approaches a periodic limit cycle. Note that for such
large system sizes, the noise amplitude is comparable to the line width.
The oscillatory behavior is very similar to the predictions by pair
approximation (see Fig.~\ref{fig:pair}). Moreover, the amplitude again
increases with $b$. To demonstrate this nice agreement, the average,
minimal and maximal frequencies of defectors are depicted
in Fig.~\ref{fig:dminmax}.
The MC data is obtained by averaging over $10^4$ MC steps per sites after
suitable relaxation times for $N=10^6$ sites.  In absence of periodic
oscillations, the difference between the minimum and maximum values
increases very slowly with $b$ until for $b=b_1^{{\text{rrg}}}=1.485(1)$ a 
Hopf-bifurcation occurs. For $b>b_1$ the extremal values quickly separate
from the average value reflecting the oscillating behavior in close
agreement with pair approximation. 

Simulations for $b > b_2^{(\text{rrg})} \simeq 1.82$ confirm another
prediction of pair approximation: in finite populations, oscillations with
increasing amplitudes are eventually stopped by the extinction of either
one of the strategies and the system ends up in the respective homogenous
absorbing state. This means if cooperators are to die out first, the
systems evolves into a state with all loners. Obviously any strategy may
disappear first but the probabilities depend on the parameters, the initial
configuration and the system size. For example, about 87\% (13\%) of
randomly initialized runs lead to a homogeneous loner (defector) state
for $b=1.8$ and $N=10^6$.

\begin{figure}
\centerline{\epsfig{file=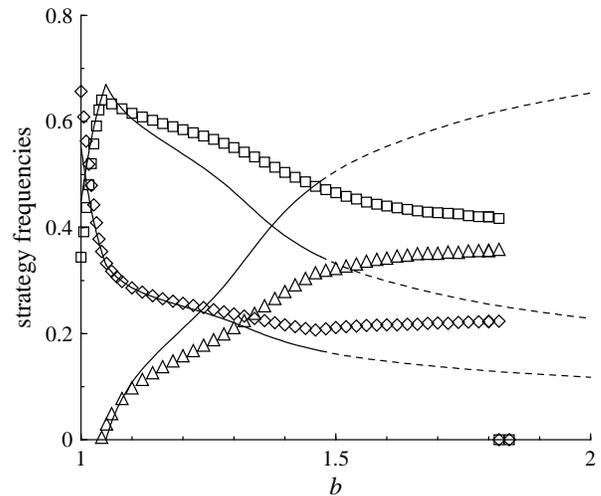,width=8cm}}
\caption{\label{fig:rrgmcp}Average frequencies of defectors (squares),
cooperators (diamonds), and loners (triangles) as a function of $b$ for
$\sigma = 0.3$ and $K=0.1$ on \RRG[]. The solid and dashed lines show the
stable and unstable predictions of pair approximation.}
\end{figure}

On \RRG the results of MC simulations are well described by pair
approximation as illustrated in Fig.~\ref{fig:rrgmcp}.
For such structures, the extinction of loners is expected to be a mean-field
type process \cite{szabo:pre00:rrg}. Rigorous analysis of our MC data
confirms that loners indeed vanish linearly with $b-b_0^{(\text{rrg})}$
where $b_0^{(\text{rrg})}=1.03858(3)$.

In the region of coexistence
($b_0^{(\text{rrg})}<b<b_2^{(\text{rrg})}$) the frequency of
loners (defectors) increases (decreases) monotonously with $b$. But note
that the cooperator frequency in MC simulations slightly increases in the
oscillating region ($b_1^{(\text{rrg})}<b<b_2^{(\text{rrg})}$).
The dashed lines in Fig.~\ref{fig:rrgmcp} represent the unstable
stationary solution of pair approximation meanwhile the symbols are
determined by averaging over more than 300 cycles in the MC simulations.

\section{\label{sec:conc}Conclusions}
The present investigations confirm that voluntary participation in PDG
interactions is capable of preventing "the tragedy of the commons" and
efficiently avoids states of mutual defection. Introducing the loner
strategy gives rise to a cyclic dominance that can maintain the co-existence
of all three strategies. With players arranged on a square lattice, the
cyclic invasions maintain a self-organizing three-color pattern. In contrast,
different types of oscillations in the frequencies of the strategies are
observed if players are located on a \RRG[]. Both scenarios have the same 
connectivity $z=4$, i.e. the players invariably interact with four neighbors.
Occasionally players reassess their strategy by sampling their neighborhood
and switch strategy with a probability depending on the payoff difference. 

\begin{figure}
\centerline{\epsfig{file=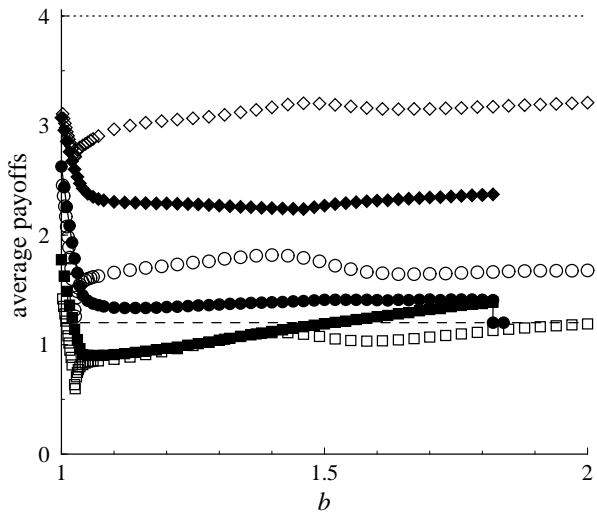,width=8cm}}
\caption{\label{fig:avpo}Average payoff of cooperators (diamonds) and
defectors (squares) as well as the average population payoff (circles) a
function of $b$ for $K=0.1$ and $\sigma=0.3$. The dashed line indicates the
loners (fixed) income $\sigma$ and the dotted line shows the maximal payoff
available for uniform cooperation. Closed (open) symbols refer to MC
simulations on \RRG (square lattice).}
\end{figure}

From the view point of sociology (or behavioral sciences) the most relevant
questions refer to the average individual income of defectors, cooperators,
and loners. Figure~\ref{fig:avpo} clearly illustrates that spatial
extension promotes cooperation. 
The average payoff of cooperators is significantly higher than that of
defectors but more importantly, the average population payoff is larger than
the loners income $\sigma$. This means that the possibility to participate
in the PDG has a positive net return for the population. At the same time
it remains far below the maximum $m_c=4$ leaving room for further strategical
improvements.

Comparing the results for square lattices and \RRG[], we note that the
average payoff of cooperators, and even more pronounced the average
population payoff, turns out to be significantly lower on \RRG[]. Due
to the randomly drawn links between the players, cluster formation becomes
more difficult while facilitating exploitation. But recall that in absence
of spatial structures, only loners survive and forego the chances of
the PDG. Similarly, for a sufficiently high temptation to defect ($b>b_2$)
the oscillations on \RRG usually terminate in the absorbing state of all
loners (see above).

In this work we investigate effects of the underlying geometry on the
success of cooperative strategies by comparing the game dynamics on square
lattices and \RRG with equal connectivity. These two structures are
indistinguishable at the level of pair approximation. However, MC
simulations reveal significant qualitative differences arising from the
distinct topologies. At the same time, pair approximation turned out to
reproduce a detailed picture of the MC results obtained on \RRG[].

\begin{acknowledgments}
G.~S. acknowledges support of the Hungarian National Research Fund under
Grant No. T-33098. Ch.~H. acknowledges support of the Swiss National Science 
Foundation.
\end{acknowledgments}


\end{document}